\documentclass[prd,twocolumn,showpacs,showkeys,nofootinbib,superscriptaddress]{revtex4}
\usepackage{graphicx}
\usepackage{dcolumn}
\usepackage{epsfig}
\usepackage{bm}

\begin{document}

\topmargin -0.50in
\title{Charged current cross section for massive cosmological neutrinos
impinging  on radioactive nuclei}

\author{R.~Lazauskas$^{1}$, P.~Vogel$^{2}$, C. Volpe}
\email{lazauskas@lpsc.in2p3.fr}
\email{pxv@caltech.edu}
\email{volpe@ipno.in2p3.fr} 
\affiliation{ 
Institut de Physique Nucl\'eaire, F-91406 Orsay cedex, France.\\
$^{2}$ Kellogg Radiation Laboratory, Caltech,
Pasadena, California, 91125, USA.}

\begin{abstract}
We discuss the cross section formula both for massless and massive neutrinos on
stable and radioactive nuclei. The latter could be of interest for the detection of
cosmological neutrinos whose observation is one of the main challenges of modern cosmology.
We analyze the signal to background ratio as a function of the ratio $m_{\nu}/\Delta$, i.e. the
neutrino mass over the detector resolution and show that an energy resolution 
$\Delta \le 0.5$~eV would be required
for sub-eV neutrino masses, independently of the gravitational neutrino clustering. 
Finally we mention the non-resonant character of neutrino capture on radioactive nuclei. 
\end{abstract}

\maketitle

\section{Introduction}
Modern big-bang cosmology firmly predicts the existence of a relic neutrino background,
and relates its temperature to the temperature of the background microwave radiation
\begin{equation}\label{eq:1}
T_{\nu}/T_{\gamma} = (4/11)^{1/3} ~,
\end{equation}
see, e.g. the basic texts \cite{Peebles,Wein_c}\footnote{
Corrections to the above ratio caused by the incomplete neutrino decoupling are only
at a few percent level \cite{mangano}.}.  Verifying the existence of the relic neutrino
sea represents one of the main challenges of modern cosmology.

Clearly, in contrast to the study of the  background microwave radiation that has
a long history and has reached an
unprecendented accuracy (see, e.g. the latest results in \cite{Spergel:2006hy}), detection of  relic
neutrinos remains an unfulfilled dream. 
Various strategies have been proposed so far, based on laboratories searches 
\cite{orpher,Lewis:1979mu,Cabibbo:1982bb,Langacker:1982ih,Stodolsky:1974aq,Hagmann:1998nz,Smith:1985fz}
and 
astrophysical observations \cite{Weiler:1982qy,Weiler:1983xx,Eberle:2004ua,Weiler:1997sh,Fargion:1997ft}, 
such as absorption dips in the flux of Ultra High Energy neutrinos (for a review see e.g. 
\cite{Gelmini:2004hg}).
As far as their detection in laboratory experiments is concerned, one needs to
overcome two main obstacles: the low cross section characteristic of weak interactions
and the low energy of relic neutrinos. The second obstacle  can be overcome if the corresponding detection reactions have vanishing thresholds. Therefore we discuss here
the possibility of detecting the relic neutrinos by the charged current reactions 
using radioactive unstable nuclei as targets. 

The paper is organized as follows:
In the next section we derive expressions for the charged current 
neutrino induced reaction cross sections involving nonrelativistic neutrinos.
We show that such cross sections, when the
corresponding reaction has a vanishing threshold,
scale with $c/v_{\nu}$ so that the number of events converges to a constant
for $v_{\nu} \rightarrow 0$. In the following section we discuss the possibility
to use a tritium target to detect the cosmological $\nu_e$. We show that the main
challenge is the separation of the produced electrons 
with energies just above the endpoint of the $\beta$ spectrum from
the overwhelming flux of the electrons from the tritium $\beta$ decay that extends just below
the 18.6 keV endpoint.  We also discuss the possibility of a resonance enhancement
of reactions involving cosmological neutrinos. We show that  the charged current
reactions, included the radiative ones, do not have a resonance character.
In the conclusion we summarize our findings and stress the need for an extreme
energy resolution if sensitivity to detect sub-eV mass relic neutrinos should be
reached.

\section{Cross sections for massive neutrinos}
Let us first recapitulate briefly the cross section for massless neutrinos.
We use the reaction 
\begin{equation}\label{e:2}
\bar{\nu}_e + p \rightarrow e^+ + n
\end{equation}
as an example. This can be easily modified for reactions without
threshold such as
\begin{equation}\label{e:3}
\nu_e + n \rightarrow e^- + p ~.
\end{equation}
\noindent
Since we are interested in very low energy neutrinos, we can treat nucleons nonrelativistically,
and keep only the lowest order terms in $E_{\nu}/M$ and $E_e/M$.
The standard expression is then ($\hbar=c=1$, see e.g. \cite{Vogel:1999zy})
\begin{equation}\label{e:4}
\frac{d \sigma}{d q^2} = \frac{G_F^2 \cos^2 \theta_C }{\pi} 
\frac{| {\cal{M}} |^2}{(s - M_p^2)^2} ~,
\end{equation}
where $q^2$ is the momentum transferred squared and $s$ is the square of the center-of-mass (CM) energy. 

Starting with the usual current $\times$ current weak interaction
\begin{equation}\label{e:5}
[\bar{u}_n ( \gamma_{\mu} f - \gamma_{\mu} \gamma_5 g ) u_p]
[\bar{v}_{\nu} \gamma^{\mu} (1 - \gamma_5) v_e]
\end{equation}
where $f,g$ are the vector and axial-vector form factors respectively,
we arrive at the squared matrix element (see, e.g. \cite{Azimov})
\begin{eqnarray}
| {\cal{M}} |^2 &=& (f + g)^2 (p_p \cdot p_e)(p_n \cdot p_{\nu})
+ (f - g)^2 (p_n \cdot p_e) \nonumber \\
&& (p_p \cdot p_{\nu}) + (g^2 - f^2) M_n M_p (p_e \cdot p_{\nu}) ~.
\label{e:6}
\end{eqnarray}
Evaluating it in the laboratory frame where the proton is at rest, and keeping only the leading terms
one gets
\begin{equation}\label{e:7}
| {\cal{M}} |^2 = M_n M_p E_{\nu} E_e [ (f^2 + 3 g^2) +
(f^2 - g^2) v_e v_{\nu} \cos \theta] ~.
\end{equation}
\noindent
with $M_n,M_p$ the neutron and proton masses.
Furthermore, using $s = (p_{\nu} + p_p)^2 = M_p^2 + 2M_p E_{\nu}$ in the laboratory frame, and
using the Jacobian
\begin{equation}\label{e:8}
\frac{d q^2}{d \cos \theta} = 2 E_{\nu} p_e ~,
\end{equation}
we obtain the usual lowest order expression \cite{Vogel:1999zy}
\begin{equation}\label{e:9}
\frac{d \sigma}{d \cos \theta} = \bar{G}^2 E_e p_e
[(f^2 + 3 g^2) + (f^2 - g^2) v_e v_{\nu} \cos \theta] ~.
\end{equation} 
with $\bar{G} = G_F \cos \theta_C/ \sqrt{2 \pi}$. 

Let us now consider the case of massive neutrinos.
In Eq.(\ref{e:4}) one should  then substitute \cite{LL}
\begin{equation}\label{e:10}
(s - M_p^2)^2 \rightarrow [s - (M_p + m_{\nu})^2][s - (M_p - m_{\nu})^2] = 4 M_p^2 p_{\nu}^2 ~,
\end{equation}
where the last expression is again in the laboratory frame. The Jacobian in Eq.(\ref{e:8}) becomes instead
\begin{equation}\label{e:11}
\frac{d q^2}{d \cos \theta} = 2 p_{\nu} p_e ~.
\end{equation}
The cross section is then given by (using $M_n \sim M_p$ as in the Eq. (\ref{e:8}))
\begin{equation}\label{e:12}
\frac{d \sigma}{d \cos \theta} =  \frac{\bar{G}^2}{v_{\nu}} E_e p_e
[  (f^2 + 3 g^2) + (f^2 - g^2) v_e v_{\nu} \cos \theta] ~.
\end{equation}
presenting a $1/v_{\nu}$ dependence\footnote{Note that in Ref.\cite{Strumia:2003zx}, where 
the cross section for  charged current neutrino capture on a free neutron, Eq.(\ref{e:3}), was evaluated, these modifications were not made and are not reflected in Fig. 1 of that paper. }. 
Such a behaviour is in agreement with the
general form for the cross section associated to exothermic reactions of nonrelativistic particles \cite{Landau}.
We see that
this form is a ``natural one" that in most cases of practical importance,
when $v_{\nu} \rightarrow c$, acquires the standard form Eq.(\ref{e:9}). 

The interaction cross section of very low energy neutrinos, 
Eq.(\ref{e:12}) was implicitly used long time ago by Weinberg \cite{Wein}. Very recently this process has attracted particular interest thanks to the work by Cocco {\it et al.} \cite{Cocco} where 
the authors have considered the possibility of detecting cosmological neutrinos 
through their capture  on radioactive beta decaying 
(and hence with no threshold) nuclei. 

In the case of nuclear (stable or unstable) targets,
i.e. reactions
\begin{equation}\label{e:13}
\nu_e + A_Z \rightarrow e^- + A_{Z+1}^* {\rm ~~or~~}  \bar{\nu}_e + A_Z \rightarrow e^+ + A_{Z-1}^*
\end{equation}
when possible, one would use the usually known $ft$ value
of the inverse radioactive decay to
 eliminate the fundamental
constants and the nuclear matrix element
 (see e.g. \cite{Bahcall})
 \begin{equation}\label{e:14}
 \sigma = \sigma_0 \times \langle \frac{c}{v_{\nu}} E_e p_e F(Z,E_e) \rangle \frac{2I'+1}{2I+1}
\end{equation}
with
 \begin{equation}\label{e:15}
 \sigma_0 = \frac{G_F^2 \cos^2 \theta_C {m_e}^2 }{\pi}| M_{nucl}|^2  = 
 \frac{2.64 \times 10^{-41}}{f t_{1/2}}. 
 \end{equation}
in units of {\rm cm}$^2$.
 Here the averaging is done over the incoming flux, $t_{1/2}$ is in seconds, the
 statistical function $f$ and the electron energy $E_e$ and
 momentum $p_e$ are evaluated with $m_e$ as a unit of energy,
 and the nuclear matrix element is excluded using
 the relation
 $| M_{nucl}|^2 \simeq  6300/ft_{1/2}$ .
 For the neutrino capture on stable targets (i.e. with a threshold) the
 electron energy is simply
 (neglecting recoil) $E_e = E_{\nu} - E_{thres} + m_e$. So, for the
 $\bar{\nu}_e$ induced reactions  $E_{e^+} = E_{\bar{\nu}} - Q_{\beta} - m_e $
 and for the $\nu_e$ induced reactions $E_{e^-} =  E_{\nu} + Q_{EC} + m_e$.
 For the
 capture on a radioactive target   $E_{e^-} = E_{\nu} + Q_{\beta} + m_e$ and
 $E_{e^+} = E_{\bar{\nu}} +Q_{EC} + m_e$.
 
 \section{Applications}

Let us now consider the possibility to use the neutrino capture  by radioactive beta-decaying nuclei to detect cosmological neutrinos. For this aim 
the relevant quantity is the number of events, i.e. the
cross section times the flux. For the latter, the dependence on $v_{\nu}$  in Eqs.(\ref{e:12} ,\ref{e:14}) 
is canceled out 
and one obtains the number of events that converges to a constant when $v_{\nu} \rightarrow 0$, in agreement with \cite{Wein,Cocco}.
As an example of the possible application of the above finding let us consider
 the $\nu_e$ capture on tritium, as done in \cite{Cocco}.
 While our conclusions are qualitatively similar to the conclusions reached by 
 Cocco {\it et al.} \cite{Cocco}, they differ in several significant details. 
  
Tritium decays into $^3$He with the half-life of 12.3 years.
 The decay $Q_\beta$ value is 18.6 keV, and $ft_{1/2}$ = 1143.
 From Eq.(\ref{e:14}-\ref{e:15}) we deduce the cross section for $T=1.9$ K 
 nonrelativistic neutrinos
 \begin{equation}\label{e:16}
 \sigma = 1.5 \times 10^{-41} \left( \frac{m_{\nu}}{{\rm eV}} \right) {\rm ~cm^2}~,
 {\rm ~~or~~}
 \sigma \frac{v_\nu}{c} \simeq 7.6 \times 10^{-45} {\rm cm}^2~.
 \end{equation}
 Here, in the first equation we used that $v_{\nu}/c \sim 3T/m_{\nu}$. In making that estimate
 we neglected the $v_{\nu} \sim 10^{-3}c$ virial motion of massive neutrinos
 in the galactic halo, and the motion of Earth and Solar System with respect to the
 random motion of the neutrinos.
 
 In order to evaluate possible count rate, we have to know the number density of the background
 $\nu_e$ sea. Its average value, for neutrinos evenly distributed throughout
 the whole Universe, corresponds to $T_{\nu} \sim 1.9$ K. 
 For neutrinos (or antineutrinos) of one flavor only 
 $\langle n_{\nu} \rangle$ is $\sim$ 55 cm$^{-3}$. Massive neutrinos
 will be gravitationally clustered on the scale of 
 $\sim$ Mpc for neutrinos with $m_{\nu} \sim$ 1 eV, that is on the scale of galaxy clusters
 (probably the clustering scale is even larger). 
 Assuming that in that case
 the ratio of the dimensionless neutrino and baryon
 densities $\Omega_{\nu}/\Omega_b \sim 0.5 \frac{m_{\nu}}{{\rm (eV)}}$ 
 remains the same as in the Universe as a whole, we obtain
 \begin{equation}\label{e:17}
 \frac{n_{\nu}}{\langle n_{\nu} \rangle} \sim 9 \times 10^6 n_b  
 \left( \frac{m_{\nu}}{{\rm eV}} \right)
 \sim 10^3 - 10^4 ~.
 \end{equation}
 for $m_{\nu} = 1$ eV and $n_b = (10^{-3} - 10^{-4})$ cm$^{-3}$ for a cluster of galaxies.
 In the following we do not use the last estimate and treat this ratio as an unknown
 $m_{\nu}$ dependent parameter.
 A more elaborated 
 study of neutrino clustering is made for example in \cite{Ringwald:2004np},
giving smaller but nonnegligible clustering for $m_{\nu}=$1 eV. 

Note, however, that much larger neutrino clustering was considered
in  Refs. \cite{Wigmans,Hwang}. The authors of these
papers speculate that features in the cosmic-ray spectra, in particular the
`knee' at $\sim$3 PeV, are associated with the threshold of the
$p + \bar{\nu}_e \rightarrow n + e^+$ reaction on $\sim$0.5 eV mass
neutrinos. The physics basis for the required 
clustering of $ \frac{n_{\nu}}{\langle n_{\nu} \rangle}
\sim 10^{13}$ is not provided in those papers.

 The capture rate per tritium atom is
 \begin{equation}\label{e:18}
 R  = \sigma \times\ v_{\nu} \times n_{\nu}
 \simeq 10^{-32} \times \frac{n_{\nu}}{\langle n_{\nu} \rangle} {~\rm s}^{-1} ~.
 \end{equation}
 Let us assume, probably much too optimistically, that one can use a Megacurie source of tritium
 (1 Mcu = 3.7$\times 10^{16}$ decays/s, i.e. 2.1$\times 10^{25}$ tritium atoms
 $\sim$ 100 g of tritium). The number 
 of events is then
 \begin{equation}\label{e:19}
 N_{\nu ~ capt} \simeq 6.5 \times  \frac{n_{\nu}}{\langle n_{\nu} \rangle} {~\rm year^{-1} Mcu^{-1}} ~.
 \end{equation}
 Thus, if our assumption about  the gravitational clustering is at least nearly
 correct, the capture rate
 would be reasonably large. However, the main issue would be whether the 
 primordial background neutrino
 capture signal would be detectable given the overwhelming rate of the radioactive decay.
 
 \begin{figure}[htb]
\begin{center}
\includegraphics[scale=0.35,angle=-90]{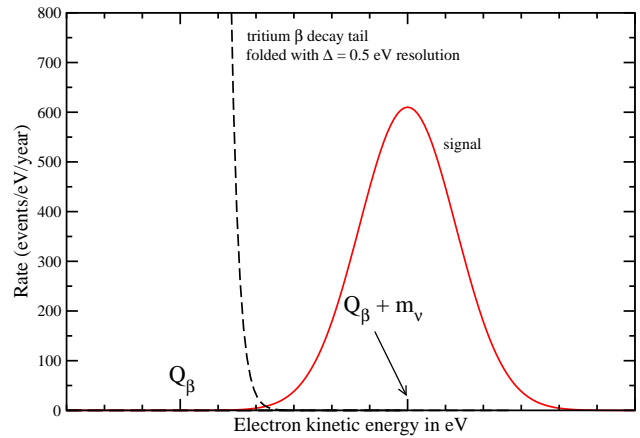}
\caption{An illustration of the spectrum of detected electrons. The 1 eV mass of the 
neutrinos is assumed, and resolution (full width at half maximum)
$\Delta$ = 0.5 eV. The tail of the tritium 
$\beta$-decay spectrum, folded with that resolution 
is depicted with the dashed curve. The signal, also folded with the 
Gaussian resolution function is shown by the full line for 
$ \frac{n_{\nu}}{\langle n_{\nu} \rangle} = 50$.}
\label{fig:schematic}
\end{center}
\end{figure}

 Electrons from the ordinary $\beta$ decay are distributed over the kinetic energy interval 
 $(0 - (Q_{\beta} - m_{\nu}))$, smeared by the resolution of the detection apparatus. 
 On the other hand, electrons from the background neutrino captures 
 are monoenergetic with the kinetic
 energy $Q_{\beta} + m_{\nu}$ again smeared by resolution. Thus, the signal to noise
 ratio will critically depend on the neutrino mass $m_{\nu}$ and on the energy resolution
 $\Delta$. Note that the fraction of electrons in an energy interval of width
 $\Delta$ just below the endpoint
 is $\sim (\Delta/Q_{\beta})^3$.
 
 To appreciate the problem we show in Fig.\ref{fig:schematic} the tail of the
 spectrum of tritium $\beta$ decay folded with a Gaussian resolution
 function and the signal of the cosmological $\nu_e$ 
 capture electrons evaluated for $ \frac{n_{\nu}}{\langle n_{\nu} \rangle} = 50$
 and clearly separated in this idealized situation from the background.
 
 Remarkably, the ratio of the background neutrino capture rate and the competing $\beta$
 decay with final electron within the resolution interval $\Delta$ just below the endpoint,
 does not depend on the corresponding $Q_{\beta}$ value (see \cite{Cocco}, their Eq.(23))
 and, naturally, on the nuclear matrix element. For $m_{\nu} < \Delta$
 the corresponding ratio is
 \begin{equation}\label{e:20}
 \frac{\lambda_{\nu}}{\lambda_{\beta}} \simeq 6 \pi^2 \frac{n_{\nu}}{\Delta^3}
 ~ \simeq 2.5 \times 10^{-11}  \times \frac{n_{\nu}}{\langle n_{\nu} \rangle} 
 \times \frac{1}{(\Delta {\rm (eV)})^3} ~.
 \end{equation}
 This appears to be a hopelessly small number.
 
 Before discussing the issues further, let us point out that other sources of background,
 for example the capture of solar $pp$ neutrinos,
 are not dangerous. The total solar $pp$ neutrino flux is
 $\sim 6 \times 10^{10} \nu_e$ cm$^{-2}$ s$^{-1}$ distributed over 420 keV
 \cite{Bahcall}.
 Thus, the flux in the lowest 10 eV is only about $10^6 \nu_e$ cm$^{-2}$ s$^{-1}$,
 which is less than 1\% of the effective flux of the primordial $\nu_e$ even for
 $m_{\nu} \ge$ 1 eV.
 
One should note that the discussed method is intresting only as long as
neutrinos are non-relativistic with v$<<$c. For higher energy neutrinos (like
thermal solar neutrinos that have an estimated flux 
of $10^8 ~-~ 10^9$/cm$^2$/sec/MeV and energies $\sim$ 1 keV \cite{Wick}) 
signal becomes well separated from
radioactive ion decay background, however the number of expected events
will be very small, since one is obliged to work with very small amount of
active material. 
The $\sim$keV mass sterile neutrinos, considered in the literature
(see, e.g. \cite{Kusenko}) are unobservable due to their extremely
small mixing with the active neutrinos.

 If one could achieve a resolution $\Delta$  that is less  than the
 neutrino mass $m_{\nu}$, the signal to background ratio would increase 
 since the $\beta$ spectrum
 ends at $Q - m_{\nu}$ while the electrons from neutrino capture have energy
 $Q + m_{\nu}$. The corresponding gain, i.e. the suppression
 of the tail of the $\beta$ decay spectrum, is estimated in \cite{Cocco}.
 For $m_{\nu} \ge \Delta$ it is
 \begin{equation}\label{e:21}
 \rho \simeq \frac{1}{\sqrt{2\pi}} e^{- 2 (m_{\nu}/\Delta)^2} ~
 \end{equation} 
 i.e., it is an extremely steep function of $m_{\nu}/\Delta$. According to such estimate
a  signal to background ratio of the order of unity could be reached if the
 ratio $m_{\nu}/\Delta \sim 3$.  Since $m_{\nu}$ remains unknown,
 an experiment with a fixed resolution $\Delta$ would be able to
 observe the background neutrino sea only if $m_{\nu}$ is large enough. 
 
\begin{figure}[htb]
\begin{center}
\includegraphics[scale=0.35,angle=-90]{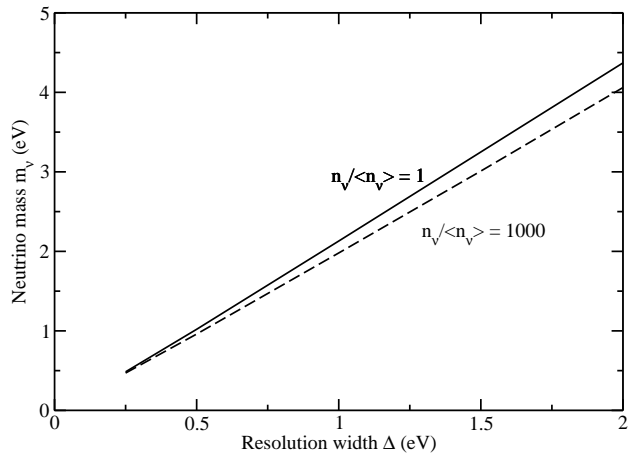}
\caption{The values of neutrino mass $m_{\nu}$ for which the signal to noise ratio
$\lambda_{\nu}/\lambda_{\beta}$ = 1 as a function of the resolution width $\Delta$.
The lines are labeled by the assumed neutrino clustering values. For details see text.}
\label{fig:ratio}
\end{center}
\end{figure}

A numerical calculation suggests that in fact $m_{\nu}/\Delta \sim 2$ is enough to achieve a
 signal to background ratio of order of unity. This is illustrated in Fig. \ref{fig:ratio} where
 $\Delta$ is the full width at half maximum of the assumed Gaussian resolution function
 and the signal as well as the background
 are centered at $Q+m_{\nu}$ and integrated over an interval of width $\Delta$.  
 The figure also
 shows that this ratio is such a steep function of $m_{\nu}/\Delta$ that it is essentially 
 independent of the enhancement of $\lambda_{\nu}$ due to the gravitational clustering.
 (Note, however, that the signal itself, i.e. the number of events, 
 is proportional to the clustering ratio
 $n_{\nu}/\langle n_{\nu} \rangle$.) Figure \ref{fig:ratio} shows that in order to achieve sensitivity to sub eV neutrino masses a resolution width below $\sim$ 0.5 eV would be necessary. Obviously, that is a very challenging requirement.
 
We wish now to discuss the possible resonance character of the neutrino capture on nuclei
at threshold. The interest in the possible resonance effects was whetted by the attempts
of Raghavan \cite {Raghavan}, that even so they are unsuccessful  so far, stimulated
lively discussion.
For the primordial background neutrinos the de Broglie wavelength is extremely large,
\begin{equation}\label{e:22}
\lambdabar_{\nu} = \frac{\hbar}{p_{\nu}} \sim 0.04 {~\rm cm}
\end{equation}
for $T_{\nu}$ = 1.9 K. 
That estimate is no longer valid for the gravitationally clustered massive neutrinos, that acquire
the corresponding virial velocity. Nevertheless, their de Broglie wavelength remains macroscopically
large.

If a resonance reaction could occur, and if it can be somehow
described by a Breit-Wigner type formula where the integrated reaction cross section
is
\begin{equation}\label{e:23}
\int \sigma_{reaction} dE = 2 \pi^2 \lambdabar^2 \frac{\Gamma_r  \Gamma_e}{\Gamma} ~,
\end{equation}
a very large cross section could be obtained. Here $\Gamma_e$ is the partial width for
the elastic scattering, $\Gamma_r$ for the capture, and $\Gamma$ is the total width.
In the case of neutrino induced reactions, obviously, there is only one channel, and 
all $\Gamma$ values should be identical.

Mikaleyan {\it et al.} \cite{reson} considered a resonance scenario 
for the endothermic reaction (not observed as yet)
\begin{equation}
\bar{\nu}_e + e^- + A_Z  \rightarrow A_{Z-1}
\label{e:24}
\end{equation}
where $A_Z$ is stable, $A_{Z-1}$ is radioactive with an endpoint $E_0$, and the 
captured electron is an orbital one in $A_Z$. The reaction occurs when $E_{\nu} = E_0 + E_b$,
where $E_b$ is the binding energy of the captured electron, so its threshold is $\sim 2m_e c^2$
below the threshold for the inverse $\beta$ decay.  
It is shown in Ref. \cite{reson} that for the process
(\ref{e:22}) the factor $\lambdabar_{\nu}^2$ in the cross section
formula Eq.(\ref{e:23})  is compensated by the $E_{\nu}^2$ dependence of the total width $\Gamma$
($\Gamma_r = \Gamma_e = \Gamma$ in this case). Moreover, the resonance electron capture cannot occur for a radioactive (hence exothermic) target $A_Z$.

While the reaction (\ref{e:24}) is indeed of resonance character and occurs only when 
the incoming $\bar{\nu}_e$ has a fixed energy,
the zero threshold reactions (with radioactive $A_Z$)
\begin{equation}\label{e:25}
\bar{\nu}_e + e^- + A_Z \rightarrow A_{Z-1} + \gamma ~~{\rm or}~~
\bar{\nu}_e + A_Z \rightarrow A_{Z-1} + e^+ 
\end{equation}
are not of resonance character. They can proceed for any energy of the incoming $\bar{\nu}_e$,
including nonrelativistic energies, but they cannot be described by the resonance formula
and the cross section is never close to  the value of $2 \pi^2\lambdabar^2$ as in Eq.(\ref{e:23}).

\section{Conclusions}

We have shown that the charged current cross section of nonrelativistic neutrinos
scales like $1/v_{\nu}$ in agreement with the generic behavior of the cross sections
for slow particles.  That means, in particular, that the charged current reactions  
with vanishing threshold  of nonrelativistic
neutrinos have a rate that converges to a finite  value as $v_{\nu} \rightarrow 0$.
With that in mind we evaluate the cross section and reaction rate of the relic
$\bar{\nu}_e$ on tritium nuclei. If a modest gravitational clustering enhancement of the relic neutrino
number density is present, the number of events is large enough that it might be
potentially observable. The more important issue is the elimination of the overwhelming
background of the electrons from tritium $\beta$ decay. We show that  signal/background ratio
of the order of unity can be achieved only if the neutrino mass $m_{\nu}$
exceeds the characteristic experimental resolution width $\Delta$ by a factor of two
or more. Thus  an energy resolution 
$\Delta \le 0.5$~eV would be required in order to be possible, even in an ideal experiment,
to detect relic neutrinos with
sub-eV neutrino masses, independently of the gravitational neutrino clustering.

\bigskip

\acknowledgements
R. Lazauskas and C. Volpe acknowledge support from the project ANR-05-JCJC-0023 "Non standard neutrino properties and their impact in astrophysics and cosmology" (NeuPAC).
P. Vogel appreciates the hospitality of the Institut de Physique Nucl\'{e}aire, Orsay and of the Aspen Center for Physics.

\end{document}